\documentclass[12pt]{article}

\usepackage{epsf,a4wide}

\def\Journal#1#2#3#4{{#1} {\bf #2}, #3 (#4)}

\def\NPB{{\em Nucl. Phys.} B}

\def\PRL{\em Phys. Rev. Lett.}
\def\PRD{{\em Phys. Rev.} D}

\def\ZPC{{\em Z. Phys.} C}
\def\PRT{\em Phys. Rep.}

\newcommand{\eqcm}{\hspace{6pt},\hspace{6pt}}

\newcommand{\X}{I}         

\begin{document}

\begin{flushright}
DAPNIA--SPHN--98--35
\end{flushright}

\begin{center}
\vskip 2\baselineskip
{\Large NON-DIAGONAL PARTON DISTRIBUTION \\[0.5\baselineskip]
FUNCTIONS\,\footnote{Talk given at the International Workshop on Deep
Inelastic Scattering and QCD (DIS~98), Brussels, Belgium}}
\vskip 2\baselineskip
{M. DIEHL} \\[0.5\baselineskip]
\textit{DAPNIA/SPhN, CEA/Saclay, 91191 Gif sur Yvette CEDEX,
France\\ E-mail: mdiehl@cea.fr}
\vskip 0.5\baselineskip
and
\vskip 0.5\baselineskip
{T. GOUSSET} \\[0.5\baselineskip]
\textit{SUBATECH\,\footnote{Unit\'e Mixte de l'Universit\'e de
Nantes, de l'Ecole des Mines de Nantes et du CNRS}, B.P. 20722, 44307
Nantes CEDEX 3, France\\ E-mail: gousset@subatech.in2p3.fr}
\vskip 2.5\baselineskip
\textbf{Abstract} \\[0.5\baselineskip]
\parbox{0.9\textwidth}{We show why the time ordering in the definition
of non-diagonal parton distributions can be omitted and discuss the
physics implications.}
\vskip 1.5\baselineskip
\end{center}

\section{The physics context}

\begin{figure}[t]
\begin{center}
  \leavevmode
  \epsfysize 2.8cm
  \epsfbox{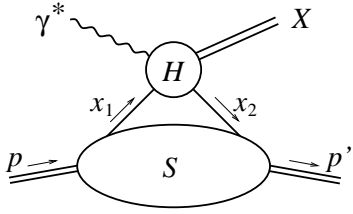}
\end{center}
\caption{\label{fig:factorise} The factorisation of $\gamma^{\ast} + p
\to X + p$ into a non-diagonal parton distribution $S$ and a hard
scattering $H$.  If $X$ is a meson, $H$ includes a distribution
amplitude describing the formation of the meson from its valence
quarks. The partons have momenta $k$ and $k'$ with plus-components
$k^+ = x_1 p^+$ and $k'^+ = x_2 p^+$.}
\vspace{-0.2cm}
\end{figure}

The idea that the gluon distribution in the proton is relevant to
diffractive photon-proton reactions~\cite{LoeRysBro} has been very
fruitful and now been applied to a variety of processes: diffractive
production of vector mesons, of a real photon or $Z$, of dijets and of
open charm. It can be generalised to the non-diffractive region, where
gluon exchange is no longer dominant and the quark distribution is
probed~\cite{JiRad,Col}. Then one can also exchange quantum numbers
and consider final states such as $p \pi^0$ or $n \pi^+$. In all cases
one exchanges a quark-antiquark or a gluon pair, attached to a parton
distribution $S$ on the proton and to a hard scattering $H$ on the
photon side as shown in Fig.~\ref{fig:factorise}; it is required that
the photon virtuality or the final state provide a hard scale while
$t$ is small. For certain processes this factorisation has been
derived in QCD~\cite{Col}.

To make the transition from $\gamma^\ast$ to $X$ possible the hadron
momenta on either side of the blob $S$ must be different, in contrast
to usual parton distributions. We thus have new quantities,
non-diagonal parton distributions, which depend on $t$ and on the
momentum fractions $x_1$ and $x_2$ of the partons. Their difference
$x_1 - x_2$ is fixed to be $\xi = [(p - p') \cdot q] /[p \cdot q]$ by
kinematics.

\section{Time ordering: why we want to drop it \ldots}

Taking for $X$ in Fig.~\ref{fig:factorise} the incoming $\gamma^\ast$
and cutting in the $s$-channel we obtain the diagram for the inclusive
$\gamma^\ast p$ cross section via the optical theorem. The blob $S$,
described now by usual parton densities, is then cut, whereas the
non-diagonal parton distributions appear in the diagram of an
\emph{amplitude} where $S$ is not cut. On a formal level ordinary
quark or gluon distributions are Fourier transformed matrix elements
of products $\bar{\psi}_\beta(0) \psi_\alpha(z) \, |_{z^2 = 0}$ or
$G_{\mu \nu}(0) \, G_{\rho \sigma}(z) \, |_{z^2 = 0}$ of field
operators, whereas the non-diagonal ones appear with \emph{time
ordered} products.

It can however be shown that, because the separation $z$ of the fields
is lightlike, one can leave out the time ordering, in other words it
is the same whether one cuts the blob $S$ or not~\cite{Us}. Before
giving the idea of the proof let us point out why this property is so
important:
\begin{list}{---}{\setlength{\itemsep}{0pt} \setlength{\parsep}{0pt}
 \setlength{\topsep}{0pt} \setlength{\partopsep}{0pt}
 \setlength{\leftmargin}{\parindent}}
\item As non-diagonal distributions can be expressed without time
ordering they become equal to the usual parton distributions in the
limit $p = p'$.
\item If the blob $S$ is cut one can show that, apart from convention
dependent phases, the corresponding parton distributions are
\emph{real valued} due to time reversal invariance.
\item Since the blob $S$ is ``already cut'' the real and imaginary
parts of the $\gamma^\ast p$ amplitude in Fig.~\ref{fig:factorise} are
obtained just from the real and imaginary parts of the hard scattering
$H$, convoluted with the same soft quantity. This is needed to obtain
a simple dispersion relation for the amplitude~\cite{Col}.
\item While the momentum fractions $x_1$, $x_2$ can a priori range
from $- \infty$ to $+ \infty$ in the loop integral of
Fig.~\ref{fig:factorise}, they are restricted to a finite interval
because the lower blob is effectively cut. This is crucial for a
parton interpretation: no parton in a hadron can be faster than the
hadron itself.
\item An ordinary parton distribution can be interpreted as the
squared amplitude for the emission of a specified parton, summed over
all configurations of spectator partons, which go across the cut of
the blob $S$. Since non-diagonal distributions are cut as well they
have a similar interpretation. Because $x_1$ and $x_2$ are different
they involve however the \emph{interference} of two different
amplitudes. In this sense they contain essentially new information on
the proton structure.
\end{list}
The question of time ordering is also relevant for higher twist parton
distributions, {\it i.e.}\ soft blobs with more than two parton
lines. For the diagonal case, $p = p'$, Jaffe has shown that time
ordering can be dropped in leading and nonleading twist
distributions~\cite{Jaf}.

\section{\ldots\ and why we are allowed to}

The idea of our proof is to use the analytic properties of scattering
amplitudes, following Landshoff and Polkinghorne~\cite{Lan}. Let us
work in the collision c.m.\ with the 3-axis along $p$ and introduce
for any vector $v$ its light cone coordinates $v^{\pm} = (v^0 \pm v^3)
/\sqrt{2}$ and its transverse part $v_T = (v^1, v^2)$. The parton
distributions can be expressed in terms of
\begin{equation}
  \int d k^- \, d^2 k_T \left. \cal{A} \, 
  \right|_{k^+ = x_1 p^+}  \eqcm
\end{equation}
where ${\cal A}$ is the proton-parton scattering amplitude depicted by
the blob $S$, given by ${\cal A} = \int d^4 z\, e^{i k \cdot z}\,
\langle p' |\, T \bar{\psi}_\beta(0) \psi_\alpha(z) \,| p \rangle$ if
for definiteness we take quark distributions. In position space the
integral over $k^-$ corresponds to setting $z^+ = 0$ in the matrix
element, and the combined integral over $k^-$ and $k_T$ puts $z$ on
the light cone. It turns out that the integration over $k^-$ is enough
for our purpose (we may thus also consider $k_T$-unintegrated
distributions).

\begin{figure}[t]
\begin{center}
  \leavevmode
  \epsfxsize 4.6in
  \epsfbox{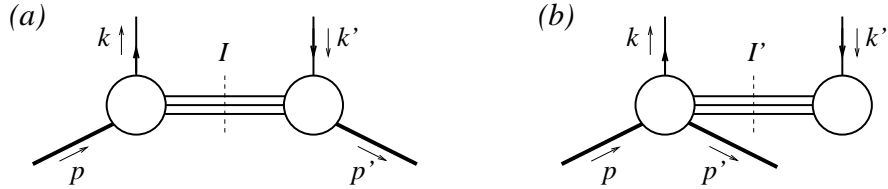}
\end{center}
\caption{\label{fig:cuts} Different ways to cut the proton-parton
amplitude ${\cal A}$: $(a)$ in the $s$-channel, $(b)$ in the parton
virtuality $k'^2$. $\X$ and $\X'$ denote the intermediate states.}
\end{figure}

According to the usual ideas on analyticity the amplitude ${\cal A}$
has singularities (poles and branch cuts) in the parton virtualities
$k^2$, $k'^2$ and in the Mandelstam variables $s = (- k + p)^2$, $u =
(k + p')^2$; in the standard convention they lie slightly below the
real axis of the variable concerned. Expressing each variable in terms
of the components of $p$, $p'$ and $k$ we can map these singularities
onto the complex $k^-$-plane. We find that their location is
controlled by the fractions $x_1$ and $x_2 = x_1 - \xi$ and have
several cases:
\begin{list}{---}{\setlength{\itemsep}{0pt} \setlength{\parsep}{0pt}
 \setlength{\topsep}{0pt} \setlength{\partopsep}{0pt}
 \setlength{\leftmargin}{\parindent}}
\item If $x_1 > 1$ or $x_1 < \xi -1$ then all singularities lie on the
same side of the real $k^-$-axis. Thus we can close the integration
contour by an infinite semicircle in the other half-plane without
encircling any singularity, and obtain a zero integral. Under the
assumption that ${\cal A}$ vanishes fast enough as $| \, k^-| \to
\infty$ at fixed $k^+$ and $k_T$ the semicircle does not contribute
and the integral along the real $k^-$-axis is zero itself. We thus
find that our distributions are only nonzero in the interval $\xi - 1
< x_1 < 1$.
\item For $\xi < x_1 < 1$ we find the singularities in $s$ above and
all others below the real $k^-$-axis. Closing the contour in the upper
half-plane our integral is given by the discontinuity of ${\cal A}$
across the $s$-cut, which according to the Cutkosky rules corresponds
to cutting the soft amplitude in the $s$-channel, as shown in
Fig.~\ref{fig:cuts} $(a)$. This cut amplitude can be written in terms
of $\langle p' |\, \bar{\psi}_\beta(0) \,| \X \rangle \, \langle \X
|\, \psi_\alpha(z) \,| p \rangle$ summed over all intermediate states
$\X$; using the closure property we find just the product
$\bar{\psi}_\beta(0) \psi_\alpha(z)$ without time ordering as we
wanted.
\item For $\xi - 1 < x_1 < 0$ we can pick up in a similar way the
$u$-channel cut of the soft amplitude and obtain a matrix element of
$- \psi_\alpha(z) \bar{\psi}_\beta(0)$. While for $\xi < x_1 < 1$ we
had the emission of a quark with momentum fraction $x_1$ and its
reabsorption with fraction $x_2$ we now have emission of an antiquark
with fraction $- x_2$ and its reabsorption with fraction $- x_1$.
\item The region $0 < x_1 < \xi$ has the particularity that $x_1$ is
positive but $x_2 $ negative; in a parton picture one has now the
transition from a proton $p$ to a proton $p'$ and a quark-antiquark
pair. In the $k^-$-plane the singularities in $s$ and $k'^2$ lie
above, those in $u$ and $k^2$ below the real axis. If we chose to pick
up the former two we must cut the soft amplitude in the $s$-channel as
before, but we must also cut in $k'^2$, as shown in
Fig.~\ref{fig:cuts} $(b)$. Altogether we find precisely the sum over
intermediate states $\X$ and $\X'$ needed to go from a time ordered
product of quark fields to an ordinary one.
\end{list}
In summary, we can show under fairly general assumptions on the
amplitude ${\cal A}$ that in parton distributions the time ordering of
fields is irrelevant because the component $z^+$ of their separation
is zero, in other words because parton distributions are integrated
over the minus component of the parton momentum.

\begin{thebibliography}{99}

\bibitem{LoeRysBro} J. Bartels and M. Loewe,
\Journal{\ZPC}{12}{263}{1982}; M.G. Ryskin,
\Journal{\ZPC}{57}{89}{1993}; S.J. Brodsky {\it et al},
\Journal{\PRD}{50}{3134}{1994}.

\bibitem{JiRad} X. Ji, \Journal{\PRL}{78}{610}{1997};
\Journal{\PRD}{55}{7114}{1997}; A.V. Radyushkin,
\Journal{\PRD}{56}{5524}{1997}.

\bibitem{Col} J.C. Collins, L. Frankfurt and M. Strikman,
\Journal{\PRD}{56}{2982}{1997}; J.C. Collins and A. Freund,
hep-ph/9801262.

\bibitem{Us} M. Diehl and T. Gousset, hep-ph/9801233, to appear in
{\em Phys. Lett.}  B.

\bibitem{Jaf} R.L. Jaffe, \Journal{\NPB}{229}{205}{1983}.

\bibitem{Lan} P.V. Landshoff and J.C. Polkinghorne,
\Journal{\PRT}{5}{1}{1972}.

\end{thebibliography}
\end{document}